\documentclass[aps,preprintnumbers, prd, floatfix,twocolumn,preprintnumbers, letterpaper, superscriptaddress,nofootinbib]{revtex4}
\usepackage[utf8]{inputenc}
\usepackage{amsmath,amssymb}
\usepackage{amsfonts}
\usepackage{amsmath}
\usepackage{amssymb,epsf}
\usepackage{latexsym}
\usepackage{graphicx,epsfig}
\usepackage{amssymb}
\usepackage{float}
\usepackage{subfigure}
\usepackage{epstopdf}
\usepackage[colorlinks=true,citecolor=blue,linkcolor=blue,urlcolor=black]{hyperref}
\usepackage{dcolumn}
\usepackage{psfrag}
\usepackage{wrapfig}
\usepackage{makeidx}
\usepackage{epsf}
\usepackage{color}
\usepackage{multirow}
\usepackage{tikz}

\begin{document}
\title{Big-Bang nucleosynthesis constraints on (dual) Kaniadakis cosmology}
\author{Ahmad Sheykhi$^{1,2}$\footnote{asheykhi@shirazu.ac.ir}, Ava Shahbazi Sooraki$^{1}$, Leila Liravi}
\address{ Department of Physics, College of
Science, Shiraz University, Shiraz 71454, Iran\\
$^{2}$ Biruni Observatory, College of Science, Shiraz University,
Shiraz 71454, Iran}

\begin{abstract}
We investigate the concept of Kaniadakis entropy and its dual
formulation, examining their implications for gravitational
dynamics within the framework where gravity emerges as an entropic
force resulting from changes in the informational content of a
physical system. In this context, we derive a modified form of
Newton's law of gravitation that reflects the corrections
introduced by both Kaniadakis entropy and its dual state.
Furthermore, we apply the emergent gravity scenario at large
scales and derive the modified Friedmann equations incorporating
corrections from (dual) Kaniadakis entropy.  Our results provide
deeper insights into the interplay between thermodynamics and
gravitational dynamics. In order to constrain the model parameter,
we study the Big-Bang Nucleosynthesis in the context of (dual)
Kaniadakis cosmology. We explore an alternative method to
establish limits on the Kaniadakis parameter, denoted as $K$, by
examining how (dual) Kaniadakis cosmology influences the
primordial abundances of light elements i.e. Helium
$_{}^{4}\textit{He}$, Deuterium $D$ and Lithium
$_{}^{7}\textit{Li}$. Our analysis indicates that the obtained
ranges for the dual Kaniadakis parameter (unlike the Kaniadakis
parameter) exhibit overlap for the aforementioned light elements,
and the allowed values fall within the range $ -0.8\times
10^{-78}\lesssim \tilde {K^*}\lesssim   0.8\times 10^{-78}$, which
shows that the deviations from the conventional Bekenstein-Hawking
formula are minimal, as expected. This consistency between the
ranges suggests a potential solution to the well-known
\textit{Lithium problem}. Furthermore, we discuss the relationship
between cosmic time $t$ and temperature $T$ within the framework
of (dual) Kaniadakis cosmology. We observe that an increase in the
Kaniadakis parameter leads to a rise in the temperature of the
early universe. Conversely, when the dual Kaniadakis parameter
increases, the temperature of the early universe decreases.
\end{abstract}
 \maketitle

 \newpage
\section{Introduction\label{Intro}}
Black hole thermodynamics suggests that gravitational field
equations, such as Einstein's general relativity, could emerge
from fundamental thermodynamic principles. This conjecture is
founded on the profound relationship between thermodynamic
quantities such as entropy and temperature and the geometric
properties of the horizon, including its area and surface gravity.
The evidence for the thermodynamic-gravity correspondence is
multifaceted. At its core is the realization that Einstein's field
equations can be derived from the fundamental thermodynamic
relation  $\delta Q=T\delta S$ when applied to local horizons.
This derivation, first achieved by Jacobson \cite{Jac},  extended
beyond Einstein's gravity to encompass $f(R)$ gravity \cite{Elin},
Gauss-Bonnet gravity, scalar-tensor gravity, and the more general
Lovelock gravity \cite{Pad1,Pad2,Cai1,Pad3}. In cosmological
contexts, this connection becomes particularly striking. In the
context of Friedmann-Robertson-Walker (FRW) cosmology, it has been
shown that the first Friedmann equation, when applied at the
apparent horizon, is directly related to the first law of
thermodynamics, expressed as $dE=TdS+WdV$. This relationship works
both ways \cite{wang1,wang2,CaiKim,Cai2,Shey11,Shey22,Shey3,Shey4}.
The connection between thermodynamic principles at the boundary
and gravitational field equations in the bulk is crucial for
understanding the holographic principle. The thermodynamic view of
gravity is further supported by ideas from statistical mechanics.
Verlinde's theory of entropic gravity \cite{Ver} demonstrated that
gravity can emerge from fundamental principles such as the
holographic principle and the equipartition law of energy,
specifically applied to the degrees of freedom at the horizon. In
this framework, variations in the system's information content
lead to the generation of an entropic force, which can be
expressed mathematically by the gravitational field equations.
This suggests that gravity may not be a fundamental force but
rather an emergent phenomenon resulting from the statistical
properties of the underlying system. The entropic nature of
gravity has been thoroughly investigated (see, for example,
\cite{Cai3,sheyECFE,Visser,SRM} and the references therein). The
most profound implication comes from Padmanabhan's work on
emergent gravity, which proposes expansion of the universe can be
understood as a consequence of space emerging itself \cite{PadEm}.
In this concept, cosmic space develops as cosmic time advances.
The core ideas in this framework are the degrees of freedom of
matter fields in both the bulk and on the boundary, with both
playing a significant role. By determining the difference in
degrees of freedom between the boundary and the bulk and equating
it to the change in volume, the Friedmann equations can be derived
\cite{CaiEm,Yang,Sheyem}.

It is essential to emphasize that, all discussed approaches share
one crucial feature: the entropy expression fundamentally shapes
how gravitational field equations emerge from thermodynamics.
Modify the entropy, and you necessarily modify the gravitational
theory
\cite{CaiLM,SheT1,Odin1,SheT2,Emm2,SheB1,SheB2,Odin2,Odin3,Odin4,Odin5}.

In the present work, by applying the (dual) Kaniadakis entropy
formalism to horizon thermodynamics, we derive and analyze the
resulting modifications to cosmological field equations. Previous
studies \cite{Lym,Shey1,Shey2} have examined the role of
Kaniadakis entropy,  in cosmological frameworks. The analysis in
\cite{Lym} explored how Kaniadakis entropy modifies cosmology by
examining the energy exchange at the universe's apparent horizon.
Their approach was based on the fundamental thermodynamic relation
$-dE=TdS $, where the $-dE$, represents the outward energy flux
across the horizon over an infinitesimal time interval. This
formulation demonstrated that the modified Friedmann equations
acquire additional terms, which can be interpreted as an effective
dark energy component influenced by the Kaniadakis parameter
\cite{Lym}. While \cite{Lym} studied entropy effects through
energy fluxes, the authors of Refs. \cite{Shey1,Shey2} took a
geometrically focused approach, modifying only the left-hand side
of the field equations. This distinction is crucial because:
entropy fundamentally originates from the geometric properties of
spacetime. Moreover, considering the universe's expansion, the
work term related to volume change should be included in the first
law of thermodynamics, formulated as  $dE=TdS+WdV$
\cite{Shey1,Shey2}. Our work differs from \cite{Lym,Shey1,Shey2}
in that they derived modified Friedmann equations using the first
law of thermodynamics with Kaniadakis entropy, while here we work
in the framework of entropic force and emergent gravity and by
incorporating both Kaniadakis and dual Kaniadakis entropy
correction terms. Our approach provides a new perspective on the
effects of generalized entropy corrections in cosmology and their
influence on gravitational dynamics.

On the other side, Big-Bang Nucleosynthesis (BBN) is a process
that occurred in the early universe shortly after the Big-Bang,
during a time when the universe was extremely hot and dense. This
process took place approximately between $0.01$ seconds and a few
hundred seconds (around $3$ minutes) after the Big-Bang. During
BBN, the high temperature enabled nuclear reactions, resulting in
the formation of light elements such as Helium
$_{}^{4}\textit{He}$, Deuterium $D$ and Lithium
$_{}^{7}\textit{Li}$. As the universe expanded and cooled, the
nuclear reactions gradually slowed and finally ceased, leading to
the observed amounts of these light elements that we observe in
the universe today. It is well-known that both BBN and the Cosmic
Microwave Background (CMB) radiation  serve as strong evidences
that our universe was once extremely hot and dense in its early
stages.

The theory of BBN imposes strict constraints on various
cosmological models, requiring consistency between theoretical
predictions and observational data (see
\cite{Sahoo1,Anish,Luciv,Sahoo2} and references therein). Modified
cosmological models can influence the primordial abundances of
these light elements \cite{Anish}. While standard BBN theory
effectively accounts for the observed amounts of helium and
hydrogen, it faces challenges with Lithium, leading to the
so-called \textit{cosmological Lithium problem}. This issue is a
significant topic of discussion in modern cosmology, and
researchers are interested in whether it can be resolved within
modified cosmological frameworks. The aim is to investigate the
cosmological Lithium problem within the context of modified
cosmology, specifically looking at how certain values of model
parameters might alleviate the issue.

Our aim in this work is to explore the implications of (dual)
Kaniadakis cosmology on BBN and derive constraints on both
Kaniadakis ($K$) and dual Kaniadakis ($K^*$) parameters. In this
context, we can evaluate the feasibility of (dual) Kaniadakis
cosmology by applying the conditions of BBN. The main difference
between this work and the work presented in the \cite{Luciv}, is
that in this paper we constrain the Kaniadakis parameter and its
dual (deformed) state using the available observational data for
the abundances of the light elements helium, deuterium, and
Lithium, and we examine the \textit{Lithium problem} within the
framework of (dual) Kaniadakis cosmology. Another key difference
in our work is that for constraining the Kaniadakis parameter and
its dual parameter, we use the Friedmann equation derived from the
emergent gravity approach, whose sign and coefficient of the
correction term differ from those obtained via the first law of
thermodynamics, which leads to different results for BBN
constraints. Additionally, we will examine the relationship
between cosmic time and temperature in the early universe,
particularly in the framework of modified (dual) Kaniadakis
cosmology, which is a topic that has not been previously addressed
in the literatures. Our analysis shows that the early universe
becomes hotter with an increase in the Kaniadakis parameter, while
an increase in the dual Kaniadakis parameter results in a decrease
in the temperature of early universe.

This paper is structured as follows. Section II presents a
comprehensive overview of (dual) Kaniadakis entropy's theoretical
foundations and its implications for black hole thermodynamics and
MOND theory. In section III, we utilize the entropic force
framework to derive the modified Friedmann equations. In section
IV, we employ the concept of emergent gravity to derive the
modified Friedmann equations. In section V, we constrain both
Kaniadakis and dual Kaniadakis parameters using BBN data within
the framework of (dual) Kaniadakis cosmology. Section VI then
establishes the time-temperature relation in this cosmological
model. Finally, we conclude with closing remarks in section VII.
\section{Modified (Dual) Kaniadakis entropy \label{Kan}}
Let us start with reviewing the origin of the Kaniadakis entropy.
Kaniadakis entropy is the generalization of
Boltzmann-Gibbs-Shannon entropy which contains one new parameter.
It has been argued that the Kaniadakis entropy can be expressed as
\cite{Kan1,Kan2,
    Abreu:2016avj,Abreu:2017fhw,Abreu:2017hiy,Abreu:2018mti,Yang:2020ria,
    Abreu:2021avp}
\begin{equation}
    \label{Sk1}
    S_{K} =-k_{_B}\sum^{W}_{i=1}\frac{P^{1+K}_{i}-P^{1-K}_{i}}{2K},
\end{equation}
where $P_i$ denotes the probability in which the system to be in a
specific microstate $i$, and $W$ represents the total number of
the system configurations. Here $K$ is called the Kaniadakis
parameter which is a dimensionless parameter ranges as $-1<K<1$,
and measures the deviation from standard statistical mechanics.
For simplicity, hereafter we set $k_{_B}=c=\hbar=1$.

One may apply the Kaniadakis entropy to the black hole
thermodynamics. For this purpose, one proposes $P_i=1/W$, and
$S_{BH}=\ln W$, where $S_{BH}=A/4G$ is the usual black hole
entropy. Thus, we have \cite{Mor}
\begin{equation} \label{W1}
    W=P_i^{-1}=\exp\left[S_{BH}\right].
\end{equation}
Substituting expression (\ref{W1}) into Eq. (\ref{Sk1}), after
some calculations, we find the Kaniadakis entropy versus black
hole entropy as,
\begin{equation} \label{SK2}
    S_{K}=\sum^{W}_{i=1}P_i
    \frac{W^{K}-W^{-K}}{2K}=\frac{1}{K}\sinh{(K S_{BH})}.
\end{equation}
If we expand expression (\ref{SK2}) for $K\ll1$, we arrive at
\begin{equation}\label{kentropy2}
    S_{K}=S_{BH}+ \frac{K^2}{6} S_{BH}^3+ {\cal{O}}(K^4).
\end{equation}
When $K\rightarrow0$, the standard entropy is restored.
Alternatively, one may assume the entropy of the black hole as
\cite{AbreuNet1}
\begin{equation} \label{SKD}
    S_{BH} =\frac{W^{K}-W^{-K}}{2K}.
\end{equation}
Solving this equation for $W$, one finds
\begin{equation} \label{W2}
    W =\left(K{S_{BH}}\pm \sqrt{1+K^2{S^2_{BH}}} \right)^{1/K}.
\end{equation}
In order to have a correct limiting BG entropy, we choose the
positive sign in the above relation. Now we define the dual
(deformed) Kaniadakis entropy as \cite{AbreuNet1,Amb}
\begin{equation} \label{SKD1}
    S^{*}_{K}=\ln W =\frac{1}{K}\ln \left(K{S_{BH}}+
    \sqrt{1+K^2{S^2_{BH}}} \right),
\end{equation}
where  $S_{BH}=A/4G$, with $A$ is the horizon area. When
$K\rightarrow 0$, one recovers $S^{*}_{K}\rightarrow S_{BH}$, as
expected.

It was shown that the dual Kaniadakis entropy (\ref{SKD1}) can
provide a theoretical origin for the MOND theory
 \cite{Amb}.
MOND theory success comes from its ability to explain galaxy
rotation curves by adjusting Newton's second law, offering a
different formulation for force as \cite{Milgrom1}
\begin{eqnarray} \label{mond}
        F = m \mu\left(\frac{a}{a_0}\right) \, a \,,
    \end{eqnarray}
where $a$ represents the usual acceleration, $a_0$ is a
characteristic constant, and $\mu(x)$ is a function behaves as
    $   (   \mu(x)\approx  1 \;\; for \;\;  x \gg 1 \,)
    $
    and
    $
    (\mu(x) \approx x \;\; for \;\; x \ll 1\,).
    $
To determine a specific functional form for $\mu(x)$ based on
entropic considerations, we begin with a thermodynamically
effective gravitational force equation as \cite{sheyECFE}
 \begin{eqnarray}
        \label{Fl}
 F_{\rm eff} = \frac{G M m}{R^2} \times 4G\frac{dS}{dA} \,,
\end{eqnarray}
In this framework, $A$ represents the area of the holographic
screen, and  $S$ characterizes its entropy. The generality of Eq.
(\ref{Fl}) means it holds for any valid entropy function. As an
illustration, by using the Bekenstein-Hawking entropy formula $
S_{BH} = {A}/{4G}$ in this Equation, the standard Newtonian
gravitational force $ F = G M m/R^2$ is recovered. It has been
shown  that when in Eq. (\ref{Fl}), dual Kaniadakis modified
entropy is adopted instead of $S$, one can derive the MOND theory
\cite{Amb}. Let first take derivative of Eq. (\ref{SKD1}) to find
\begin{eqnarray}
        \label{dsa}
        \frac{dS_K^*}{d A} = \frac{  dS_{BH}  }{   dA }  \frac{ 1}{  \sqrt{ 1 + K^2 S_{BH}^2  }      } \,.
\end{eqnarray}
Where $ S_{BH} = {A}/{4G}$. Combining with Eq. (\ref{Fl}), we find
the effective gravitational force as
\begin{eqnarray}
        \label{fmond}
 F_{\rm eff} &=& \frac{G M m}{R^2}\frac{1} { \sqrt{ 1 + \left(   \frac{K \pi R^2}{ G}   \right)^2 } }
 \nonumber\\
 &=& \frac{ m \, a}{  \sqrt{ 1 + ( \frac{ a_0}{  a}    )^2 } } ,
    \end{eqnarray}
where $ a \equiv {G M}/{R^2} $ and $a_0 \equiv  K \pi  M $. Based
on this equation, the interpolating function $\mu (x)$ is realized
as
    \begin{eqnarray}
        \label{inter}
\mu \left( \frac{a }{a_0} \right) =\frac{1} {\sqrt{ 1 +  \left(
\frac{a_0}{   a } \right) ^2} } = \frac{    a  }{a_0} \left( 1+
\frac{   a ^2}{a_0^2} \right)^{- 1/2} \; ,
    \end{eqnarray}
which implies $ \mu (x)= x \left( 1+ x ^2 \right)^{- 1/2}$.
Clearly, this interpolating function satisfies the conditions for
$\mu(x)$. Therefore, when dual Kaniadakis modified entropy given
in Eq. (\ref{SKD1}) is combined with the effective force expression
Eq.  (\ref{Fl}),  the resulting model mirrors MOND theory including
its distinctive transition function Eq. (\ref{inter}), which is
precisely the standard interpolating function used in MOND theory
\cite{Begeman,Gentile}.

Note that expression (\ref{SKD1}) can be also rewritten as
    \begin{equation} \label{SKD2}
        S^{*}_{K}=\frac{1}{K}\sinh^{-1}{(K S_{BH})}.
    \end{equation}
    If we expand expression (\ref{SKD2}) for $K\ll1$, we arrive at
    \begin{equation}\label{SKD3}
        S^{*}_{K}=S_{BH}-\frac{K^2}{6} S_{BH}^3+ {\cal{O}}(K^4).
    \end{equation}
The first term is the usual area law of black hole entropy, while
the second term is the leading order Kaniadakis correction term.
It is clear that the leading order term for dual (deformed)
Kaniadakis entropy has a negative sign.

In summary, for (dual) Kaniadakis entropy, the modified entropy
expression, up to the first order correction term, can be written
as
\begin{equation}\label{SKDG}
        S_{K}=S_{BH}\pm\frac{K^2}{6} S_{BH}^3+ {\cal{O}}(K^4),
    \end{equation}
where $+$ and $-$ stands, respectively, for the Kaniadakis and
    dual Kaniadakis entropy and we have dropped the star for dual
    Kaniadakis entropy.
\section{Kaniadakis entropic corrections to Newton's law and Friedmann equations\label{Newton}}
This section explores entropic gravity principles, using entropy
expression (\ref{SKDG}) to develop correction terms for Newton's
gravitational law and modifications to the Friedmann equations.
According to Verlinde's framework \cite{Ver}, gravitational
effects arise as material bodies alter information entropy due to
their relative motion with respect to holographic screens. In this
framework, the entropic force satisfies the relation
\begin{equation}\label{eq1}
    F\triangle x=T \triangle S,
\end{equation}
for test particle displacements. Here, $\triangle x$ represents
the displacement of a particle from the holographic screen, while
$T$  denotes the temperature and $\triangle S $ corresponds to the
change in entropy on the screen. We study a configuration where a
spherical mass $M$ is enclosed by a surface $\mathcal {S}$, and a
test particle of mass $m$ is placed adjacent to  $\mathcal {S}$,
at a distance smaller than its reduced Compton wavelength
$\lambda_m={1}/{m}$, ensuring that $\mathcal {S}$ remains the
boundary separating the two masses. For a test mass $m$ displaced
by $\Delta x=\eta \lambda_m$ from the surface $\mathcal{S}$, the
variation of entropy (\ref{SKDG}) takes the form
\begin{equation}\label{eq2}
    \triangle S_{K}=\frac{\partial S_{K}}{\partial A}\triangle A
    =\left( \frac{1}{4G}\pm\frac{K^2 \pi^2}{8G^3}R^4\right) \triangle
    A.
\end{equation}
The area of the surface $\mathcal {S}$ is given by $A=4\pi R^2$,
and the total energy enclosed within this surface is $E=M$. The
surface $\mathcal {S}$ contains discrete information units "bytes"
whose quantity scales with surface area according to $A=QN$, where
$N$ is the total byte count and $Q$ is a fundamental constant.
Since $\triangle N=1$, any change in the area satisfies $\triangle
A=Q$. According to the equipartition law of energy, the
temperature  $T$ is related to the total energy on the surface
$\mathcal {S}$ is given by
\begin{equation}\label{eq3a}
    T=\frac{2M}{N}.
\end{equation}
Where we have set $k_B=1$. Using relation $N=A/Q$, we can rewrite
Eq. (\ref{eq3a}) as
\begin{equation}\label{eq3}
 T=\frac{MQ}{2\pi  R^2}.
\end{equation}
By substituting Eqs. (\ref{eq2}) and (\ref{eq3}) into
Eq. (\ref{eq1}), we obtain
\begin{equation}\label{eq4}
    F=-\frac{GMm}{{R}^2}\left(\frac{Q^2}{8\pi
        \eta  G^2}\right)\left\lbrace 1\pm\alpha R^4\right\rbrace,
\end{equation}
where $\alpha=\frac{K^2 \pi^2}{2G^2}$. We can express the modified
Newton's law of gravity by defining $Q^2=8\pi  \eta G^2$. The
result is
\begin{equation}\label{eq6}
    F=-\frac{GMm}{{R}^2}\left\lbrace 1\pm\alpha R^4\right\rbrace.
\end{equation}
This is the modified Newton's law of gravitation inspired by
(dual) Kaniadakis corrections to entropy. Note that  here $+$ and
$-$ corresponds to the Kaniadakis and dual Kaniadakis modified
entropy, respectively. In the limiting case where $K\rightarrow
0$, the parameter $\alpha$ vanishes, and the standard Newton's law
of gravitation is recovered.

Now we are going to derive the correction to the Friedmann
equations. For this purpose, we consider a universe that is
spatially homogeneous and isotropic, described by the line
elements
\begin{equation}
    ds^2={h}_{\mu \nu}dx^{\mu} dx^{\nu}+R^2(d\theta^2+\sin^2\theta
    d\phi^2),
\end{equation}
where $R=a(t)r$, $x^0=t$ and $x^1=r$. The two-dimensional metric
is given by $h_{\mu \nu}$=diag $(-1, a^2/(1-kr^2))$, where  $k =
-1,0, 1$ corresponds to open, flat, and closed universes,
respectively. The apparent horizon, which serves as a natural
boundary based on thermodynamic considerations, has a radius given
by \cite{Cai2}
\begin{equation}
    \label{radius}
    R=a(t)r=\frac{1}{\sqrt{H^2+k/a^2}},
\end{equation}
where the Hubble parameter is given by $H=\dot{a}/a$. We model the
matter and energy content of the universe as a perfect fluid,
described by the energy-momentum tensor,
$T_{\mu\nu}=(\rho+p)u_{\mu}u_{\nu}+pg_{\mu\nu},$ where $\rho$ is
the energy density and $p$ is the pressure. Since there is no
observed energy exchange between the universe and any external
environment, the local conservation of energy-momentum must hold
$\nabla_{\mu}T^{\mu\nu}=0$. This condition guarantees that the
total energy and momentum of the cosmic fluid remain dynamically
conserved and leads to the continuity equation as
\begin{equation}\label{Cont}
    \dot{\rho}+3H(\rho+p)=0.
\end{equation}
By applying Newton's second law to a test particle $m$ near the
surface $\mathcal {S}$ and incorporating the gravitational force
given by Eq. (\ref{eq6}), we arrive at
\begin{equation}\label{eq7}
    F=m\ddot{R}=m\ddot{a}r=-\frac{GMm}{{R}^2}\left\lbrace 1\pm\alpha R^4\right\rbrace.
\end{equation}
The matter inside the spherical volume $V=\frac{4}{3} \pi a^3
r^3$, has an energy density given by $\rho=M/V$. Therefore, we
have  $M=\frac{4}{3} \pi a^3 r^3\rho$. By using this relation and
$R=a(t)r$,  Eq. (\ref{eq7}) can be reformulated as
\begin{equation}\label{F72}
    \frac{\ddot{a}}{a}=-\frac{4\pi G
    }{3}\rho \left\lbrace 1\pm\alpha R^4\right\rbrace.
\end{equation}
This is the modified dynamical equation for Newtonian cosmology,
incorporating the Kaniadakis (dual Kaniadakis) entropy correction.
From another perspective, we define the active gravitational mass
$\mathcal{M}$ following \cite{Cai3} as
\begin{equation} \label{u1}
    \mathcal{M}= 2 \int_{v} dV (T_{\mu\nu}-\frac{1}{2}T g_{\mu\nu})u^{\mu} u^{\nu}= (\rho + 3p)\,\frac{4\pi}{3}a^3 r^3.
\end{equation}
Now, in the expression for $\rho$, $M$ is replaced by $\mathcal M$
in Eq. (\ref{F72}). The result is
\begin{equation}\label{addot2}
    \frac{\ddot{a}}{a} =-\frac{4\pi G
    }{3}(\rho+3p)\left\lbrace 1\pm\alpha R^4\right\rbrace.
\end{equation}
If we multiply both sides of Eq. (\ref{addot2}) by factor
$2\dot{a}a$ and applying the continuity equation (\ref{Cont}), we
obtain the result after integration
\begin{equation}\label{z1}
    {\dot{a}}^2 +k =\frac{8\pi G}{3}\int d(\rho a^2) \left\lbrace 1\pm\alpha R^4\right\rbrace.
\end{equation}
Here,  $k$ is an integration constant that can be interpreted as
the curvature parameter. To evaluate the integral explicitly, we
first solve the continuity equation (\ref{Cont}), which has a
solution
\begin{equation}\label{aa11}
    \rho ={\rho} _0 a^{-3(1+\omega)}.
\end{equation}
Where $w=p/\rho$ represents the equation of state parameter, and
$\rho_0$ denotes the present energy density. Substituting
Eq. (\ref{aa11}) into Eq. (\ref{z1}) and performing the necessary
calculations, we arrive at the expression
\begin{equation} \label{bb}
    H^2+\frac{k}{a^2} =\frac{8\pi G}{3}\rho
    \left[ 1\mp\beta( H^2+\frac{k}{a^2})^{-2}\right],
\end{equation}
where $\beta\equiv(\frac{1+3\omega}{-3\omega+3}) \alpha= (
\frac{1+3\omega}{-3\omega+3})(\frac{K^2 \pi^2}{2G^2})$ and it is
obtained based on Eq. (\ref{radius}). It is convenient to rewrite
the above equation in a simplified form as
\begin{equation} \label{bkb}
    H^2+\frac{k}{a^2} \pm\beta   \left(H^2+\frac{k}{a^2}\right)^{-1}=\frac{8\pi G}{3}\rho.
\end{equation}
where the $+$ and $-$, respectively, correspond to Kaniadakis and
dual Kanidakis entropy. In the limiting case $\beta=0$ the
standard Friedmann equation is recovered.
\section{Corrections to Friedmann equations from emergence scenario \label{Emergence}}
This section applies Padmanabhan's gravity emergence paradigm
\cite{PadEm} to derive modified Friedmann equations, incorporating
the adjusted entropy relation from Eq. (\ref{SKDG}). As previously
mentioned in Padmanabhan's proposal, the expansion of the universe
is driven by the difference between degrees of freedom on the
horizon and in the bulk. This idea will be further explored in the
following sections.  He expressed this idea mathematically as
\cite{PadEm}
\begin{equation} \label{dV}
    \frac{dV}{dt}=G(N_{\mathrm{sur}}-N_{\mathrm{bulk}}),
\end{equation}
where  $N_{\mathrm{sur}}$ and $N_{\mathrm{bulk}}$ represent the
degrees of freedom on the boundary and in the bulk, respectively.
In the case of a nonflat universe, this relation needs to be
generalized, as suggested in \cite{Sheyem}, leading to
\begin{equation}
    \frac{dV}{dt}=GRH \left(N_{\mathrm{sur}}-N_{\mathrm{bulk}}\right),
    \label{dV1}
\end{equation}
where $R$ denotes the apparent horizon radius based on Eq.
(\ref{radius}). The temperature corresponding to the apparent
horizon is assumed to be
\begin{equation}\label{T2}
    T=\frac{1}{2\pi R}.
\end{equation}
To express the number of degrees of freedom on the horizon, we use
relation $N_{\mathrm{sur}}\approx 4S_h$ in which $S_h$ represents
the  entropy of the horizon. Thus, considering Eq. (\ref{SKDG}),
we can identify
\begin{equation} \label{Nsur}
    N_{\mathrm{sur}}=4\left\lbrace \frac{A}{4G}\left( 1\pm\frac{K^2}{6}(\frac{A}{4G})^2\right) \right\rbrace.
\end{equation}
Where we have considered $A=4 \pi R^2$ as the boundary area. Thus,
we have
\begin{equation} \label{Nsur11}
    N_{\mathrm{sur}}=\dfrac{4\pi R^2}{G}\pm\gamma\dfrac{4\pi R^6}{G},
\end{equation}
where $\gamma=\frac{K^2 \pi^2}{6G^2}$ and the positive sign
corresponds to the entropy correction inspired by Kaniadakis,
while the negative sign arises from the dual formulation. The
total energy within the apparent horizon is expressed as the Komar
energy, given by
\begin{equation}
    E_{\mathrm{Komar}}=|(\rho +3p)|V, \label{Komar}
\end{equation}
where  $ V=4 \pi R^3/3$ represents the volume of the sphere
enclosed by the apparent horizon. Using the energy equipartition
theorem, one derives the count of degrees of freedom for the bulk
matter field as
\begin{equation}
    N_{\mathrm{bulk}}=\frac{2|E_{\mathrm{Komar}}|}{T}. \label{Nbulk11}
\end{equation}
This equation can be written as
\begin{equation}
    N_{\mathrm{bulk}}=-\frac{16\pi^2
    }{3}(\rho+3p)R^4,\label{Nbulk1}
\end{equation}
assuming that $\rho+3p<0$ in an expanding universe. By merging
relations (\ref{Nsur11}) and (\ref{Nbulk1}) with Eq. (\ref{dV1}),
we arrive at
\begin{equation}
    4\pi R^2 \dot{R}H^{-1}=
    G R\left( \dfrac{4\pi R^2}{G}\pm\gamma\dfrac{4\pi R^6}{G}
    +\frac{16\pi^2}{3}(\rho+3p)R^4\right).
\end{equation}
After simplifying, we have
\begin{equation}
    \frac{R^2 \dot{R}}{H}-R^3=\pm\gamma R^7+\frac{4\pi
    G}{3}(\rho+3p)R^5.
\end{equation}
Now, by multiplying both sides by factor $-\dot{a}a R^{-5}$ and
using the continuity equation (\ref{Cont}), we find
\begin{equation}
    \frac{d}{dt}\left(a^2
    R^{-2}\right)\pm\frac{d}{dt}(a^2)\gamma R^{2} =\frac{8\pi G
    }{3}\frac{d}{dt}(\rho a^2). \label{Fr1}
\end{equation}
We can perform the integration of the above equation, which leads
to the following result
\begin{equation}
    R^{-2}\pm\frac{\gamma}{2}R^{2} =\frac{8\pi G
    }{3}\rho.
\end{equation}
By defining $\frac{\gamma}{2}=\lambda$ and using Eq. (\ref{radius}), we arrive at
\begin{equation} \label{bb}
    (H^2+\frac{k}{a^2}) \pm\lambda  (H^2+\frac{k}{a^2})^{-1}=\frac{8\pi
    G}{3}\rho,
\end{equation}
where $\lambda=\frac{K^2 \pi^2}{12G^2}$. This result matches the
one derived from the entropic force in Eq. (\ref{bkb}). Note that
the $+$ is obtained by considering Kaniadakis entropy, and the $-$
is obtained from its dual entropy. In the limit $\lambda=0$, the
Friedmann equation returns to its standard form.
\section{Constraints on Kaniadakis and Dual Kaniadakis parameters from BBN \label{BBN}}
\subsection{Big-Bang nucleosynthesis in (dual) Kaniadakis cosmology}
\qquad In this section we first analyze the Kaniadakis cosmology
during the radiation dominated epoch, and after that explore the
BBN in the framework of (dual) Kaniadakis cosmology. Using Eq.
\eqref{bb}, the modified Friedmann equation inspired by (dual)
Kaniadakis entropy in the flat universe ($k=0$) can be written as
\begin{eqnarray} \label {frst}
    H^2\pm\lambda H^{-2}=\frac{8\pi G}{3}(\rho+\rho_{\Lambda}),
\end{eqnarray}
By defining $H^2 \equiv \chi $ and multiplying Eq. \eqref{frst} by
$H^{2} $ , the above equation can be rewritten as follows
\begin{eqnarray}
    \chi^2-\frac{8\pi G}{3}\rho \chi \pm\lambda=0.
\end{eqnarray}
Solving the above equation for $\chi$ and substituting $\chi$ in
terms of $H$, the following expression for the Hubble parameter is
obtained
\begin{eqnarray}\label {firstm}
    H^2=\frac{1}{2}\left( {H_{GR}^2+ \sqrt{H_{GR}^4\pm 4\lambda}} \right),
\end{eqnarray}
where $-$ and $+$ sign refer to the Kaniadakis entropy and the
dual Kaniadakis entropy, respectively, while $H_{GR}=\sqrt{\frac{8
\pi G }{3} \rho(T)}$ stands for the Hubble function in the
standard cosmology. Since $K$ is small enough, we can expand the
above expression and obtain the modified Hubble parameter as
\begin{eqnarray} \label{hk}
    H^2=H_{GR}^2\pm \frac{\lambda}{H_{GR}^2} \,\, \, \to \,\,\, H=\left( H_{GR}^2\pm \frac{\lambda}{H_{GR}^2} \right)^{1/2}.
\end{eqnarray}
We can rewrite the modified Hubble parameter (\ref {hk}) in the form
\begin{eqnarray} \label{hz}
    H(T)\equiv Z(T) H_{GR}(T),
\end{eqnarray}
where the amplification factor $Z(T)$ is defined as
\begin{align}\label{amp}
    & Z(T)\equiv \frac{H}{H_{GR}(T)}  \rightarrow \notag \\
    &   Z(T)=\frac{1}{H_{GR}}\left( H_{GR}^2\pm\frac{\lambda}{H_{GR}^2} \right)^{1/2}=1\pm \frac{\lambda}{2H_{GR}^4}.
\end{align}
Substituting the energy density of relativistic particles,
expressed as $ \rho(T)=\frac{\pi^2}{30}g_{*}T^4$ ( $g_{*}\sim10$
is the effective number of degrees of freedom and $T$ is the
temperature), $H_{GR}$, reduced Planck mass $M_p=(8\pi G)^{-1/2}
\simeq 2.4 \times 10^{18}GeV $ \cite{Luciv}, and $\lambda=\frac
{K^2\pi ^2}{12G^2}$ into Eq. (\ref {amp}), we determine the
amplification factor $Z(T )$ as
\begin{align}\label{zt}
    Z(T)=1\pm\frac{2025}{384\pi^4}\frac{K^2}{g_{*}^2G^4T^8},
\end{align}
where the $-$ sign arises from the Kaniadakis cosmology, while the
$+$ sign comes from its corresponding dual cosmology. Consistent
with expectations, when $K=0$, the amplification factor reduces to
$Z(T)=1$, and the general relativity (GR) limit is recovered.\\
\subsection*{Discussion on the dimensions of Kaniadakis Parameter}
{Let us note that the Kaniadakis parameter ($K$) is dimensionless
in the units where \(k_{_B} = c = \hbar = 1\). However, in
conventional SI units, it possesses dimensions of inverse entropy,
i.e., Kelvin per Joule ($K/J$). We will elaborate
on this point below:\\
The parameter \(K\) is indeed dimensionless, which arises from the
specific unit system employed in our derivation, where we set
\(k_{_B} = c = \hbar = 1\), and from the structure of the entropy
expansion.} {Within this unit system, the Bekenstein-Hawking
entropy is given by
\[
S_{\text{BH}} = \frac{A k_B c^3}{4 G \hbar} = \frac{A}{4G}.
\]
In this framework, the gravitational constant \(G\) is not
dimensionless. Its dimension is absorbed into the definition of
the Planck length \((\ell_{\text{Pl}}^2 = G\hbar/c^3)\). Since we
set \(\hbar = c = 1\), \(G\) inherits the dimension of length
squared (\([L]^2\)). The area \(A\) also has dimensions \([L]^2\).
Consequently, the ratio \(A/(4G)\) and thus \(S_{\text{BH}}\) is
dimensionless.} {The first-order Kaniadakis entropy correction is
expressed as Eq. \eqref{SKDG} For this equation to be
dimensionally consistent, and given that \(S_{\text{BH}}\) is
dimensionless, the term \(\frac{K^2}{6} S_{\text{BH}}^3\) must
also be dimensionless. This necessitates that \(K^2\) (and
consequently the parameter \(K\)) must be dimensionless.
Furthermore, according to Eq. \eqref{zt}, the Kaniadakis parameter is dimensionless. This is because the equation is written in the units \(k_{_B} = c = \hbar = 1\), and the coefficient $G^4T^8$ in the denominator of the fraction is consequently dimensionless.\\
However, it is important to note that in conventional units (e.g.,
SI or cgs), the Kaniadakis entropy expansion given by Eq.
(\ref{SKDG}), where \(S_{\text{BH}}\) has dimensions of
Boltzmann's constant \(k_B\) (\([S_{\text{BH}}] = [k_B] = \text{J}
\cdot \text{K}^{-1}\)). For dimensional consistency, the term
\(\frac{K^2}{6} S_{\text{BH}}^3\) must also have dimensions of
\(k_B\). Therefore:
\[
[K^2] \cdot [S_{\text{BH}}]^3 = [k_B], \implies [K^2] \cdot [k_B]^3 = [k_B], \\
\]
\[ \implies [K^2] = [k_B]^{-2}.\]
Thus, in conventional units, the Kaniadakis parameter \(K\) has
dimensions of $[K] = [k_B]^{-1} = \text{J}^{-1} \cdot \text{K}$.
We shall consider this fact in section IV where we explore Time-
Temperature relation in (dual) Kaniadakis cosmology.}
\subsection{Primordial light elements{ \textit{ $_{}^{4}\textrm{He}$, D} and \textit {Li} } in (dual) Kaniadakis cosmology}

Next we explore constraints on the Kaniadakis parameter $K$ and
dual Kaniadakis parameter, $K^*$, by examining how (dual)
Kaniadakis cosmology influences the primordial abundances of light
elements, specifically Deuterium $_{}^{2}\textit{H}$, Tritium
$_{}^{7}\textit{Li}$, and Helium $_{}^{4}\textit{He}$. By
comparing theoretical predictions with observational data on these
elements abundances, we derive bounds on $K$ and $K^* $. The main
idea is to replace the modified amplification factor $Z(T)$ found
in \eqref{zt} instead of the standard Z-factor that is associated
with the effective number of neutrino species in \cite {Luciano}.
In the standard cosmological model, the parameter $Z$ is equal to
unity. However, if we consider modifications to gravity or the
introduction of additional light particles, such as neutrinos, $Z$
may differ from unity. In such scenarios, the modified Z-factor
would be represented in the form \cite {Anish,Luciano, Boran}
\begin{align}
Z_{\nu}=\left [ 1+\frac{7}{43} (N_{\nu}-3)\right ]^{1/2},
\end{align}
where in this context, $N_{\nu} $ represents the number of
neutrino species. The baryon-antibaryon asymmetry $\eta_ {10}$,
plays an important role in our analysis \cite {Adv,Simha}. Since
we aim to focus on the effects of Kaniadakis cosmology on BBN, we
set $N_{\nu}=3 $, ensuring that any deviation of $Z$ from unity
arises solely from the effects of Kaniadakis cosmology on BBN, not
from additional particle degrees of freedom.

In the subsequent analysis, we adopt the approach outlined in
\cite{Anish, Sahoo} and summarize its key aspects as:

(i)\textit {$_{}^{4}\textrm{He}$ abundance}-the production of
$_{}^{4}\textit{He}$ begins when a proton and neutron combine to
form $_{}^{2}\textit{H}$. Subsequently, Deuterium transforms into
Tritium and  $_{}^{3}\textit{He}$ through the following processes
\begin{align}
    n + p &\rightarrow \, ^2\text{H} + \gamma \label{eq:32}, \\
    ^2\text{H} + ^2\text{H} &\rightarrow \, ^3\text{He} + n, \label{eq:33} \\
    ^2\text{H} + ^2\text{H} &\rightarrow \, ^3\text{H} + p.
\end{align}
The final production of  Helium $_{}^{4}\textit{He}$ occurs
through the following nuclear reactions
\begin{align}
    ^2\text{H} + ^3\text{H} &\rightarrow \, ^4\text{He} + n  , &
    ^2\text{H} + ^3\text{He} &\rightarrow \, ^4\text{He} + p .\label{eq:35}
\end{align}
The numerical best-fit analysis yields the following constraint
for the primordial $_{}^{4}\textit{He}$  abundance
\cite{Kneller,Annu}
\begin{equation} \label{bestfit}
    Y_p = 0.2485 \pm 0.0006 + 0.0016 \left[ (\eta_{10} - 6) + 100 (Z - 1) \right].
\end{equation}
In our analysis, $Z$ follows from Eq. (\ref{zt}), with the baryon
density parameter $\eta_ {10}$ defined as \cite{Adv,Simha}

\begin{equation} \label {et}
    \eta_{10} \equiv 10^{10}\eta_B \equiv 10^{10} \frac{n_B}  {n_{\gamma}} \simeq 6,
\end{equation}
where the baryon to photon ratio is defined as $\eta_B \equiv{n_B}
/{n_{\gamma}}$ \cite {Wamp}. By setting $Z=1$, we arrive at the
standard prediction from BBN for the abundance of
$_{}^{4}\textit{He}$, which is approximatively
$(Y_p)|_{GR}=0.2485\pm 0.0006$. Additionally, the observational
data regarding the abundance of  $_{}^{4}\textit{He}$, along with
the assumption that $\eta_ {10}= 6$ suggests that $Y_p=0.2449\pm
0.004$ \cite {Brain}. Consistency between previously mentioned
value of $Y_p$ and Eq. (\ref{bestfit}) indicates

\begin{equation}
    0.2449 \pm 0.0040 = 0.2485 \pm 0.0006 + 0.0016 \left[ 100(Z - 1) \right].
\end{equation}
So we can define the value of $Z$ as

\begin{equation} \label {zhe}
    Z = 1.0475 \pm 0.105.
\end{equation}
(ii) \textit {$_{}^{2}\textit{H}$ abundance}- Deuterium forms via
the process $ n + p \rightarrow \, ^2\text{H} + \gamma $. In a
similar manner, the Deuterium abundance can be calculated from the
numerical best fit of \cite{Adv}. It gives
\begin{equation} \label{zde}
   Y_{Dp} =10^5 \left( \frac{\mathrm{D}}{\mathrm{H}} \right)_p=
2.6(1 \pm 0.06) \left(\frac{6} {\eta_{10} - 6(Z - 1)}
\right)^{1.6}.
\end{equation}
As in the previous case, setting  $Z=1$ and $\eta_ {10}=6 $
reproduces the standard GR result $Y_{D_p}|_{GR}=2.6\pm 0.16$ . By
equating the observed Deuterium abundance constraint $
Y_{D_p}=2.55\pm 0.03$ \cite{Brain}, with Eq. \eqref {zde}, we
derive
\begin{equation}
    2.55 \pm 0.03 = 2.6(1 \pm 0.06) \left(\frac{6} {\eta_{10} - 6(Z - 1)} \right)^{1.6}.
\end{equation}
Consequently, the parameter $Z$ is constrained to
\begin{equation} \label {zobs}
    Z = 1.062 \pm 0.444,
\end{equation}
which exhibits partial overlap with the constraint from abundance
of $_{}^{4}\textit{He}$ in relation (\ref{zhe}).

(iii) \textit{$_{}^{7}\textit{Li}$ abundance} - the parameter
$\eta_ {10}$ given in Eq. (\ref{et}) successfully reproduces the
observed abundances of  $_{}^{4}\textit{He}$, $D$ and other light
elements, yet it exhibits a tension with the measured
$_{}^{7}\textit{Li}$ abundance. In fact, the predicted abundance
ratio of $_{}^{7}\textit{Li}$ in the standard cosmological model
compared to the observed Lithium abundance falls within the
following range \cite {Boran}
\begin{align*}
    \frac{\text{Li}|_{GR}}
    {\text{Li}|_{obs}}
    \in [2.4 - 4.3].
\end{align*}
Quite unexpectedly, no current nucleosynthesis scenario, including
standard BBN or its modifications, can explain low observed
abundance of  $_{}^{7}\textit{Li}$. This discrepancy is commonly
known as \textit {cosmological Lithium problem} \cite{Boran}. By
imposing consistency between observational values of Lithium
abundance $Y_{Li}=1.6\pm0.3$ \cite{Brain}, and the following
numerical best fit for the abundance of $_{}^{7}\textit{Li}$
\begin{equation}
Y_{Li} =  10^{10} \left( \frac{\mathrm{Li}}{\mathrm{H}} \right)_p
=4.82 (1 \pm 0.1)\left[\frac{\eta_{10} - 3(Z-1)} {6} \right]^{2},
\end{equation}
we can derive $Z$ as
\begin{equation} \label {lit}
Z= 1.960025 \pm 0.076675.
\end{equation}
This restriction does not coincide with the findings regarding the
abundance of $_{}^{2}\textit{H}$, in Eq. (\ref{zobs}), and
$_{}^{4}\textit{He}$ abundance, in Eq.(\ref {zhe}).
\subsection{Discussion in the results}
\begin{figure} [H]
\includegraphics[scale=0.88]{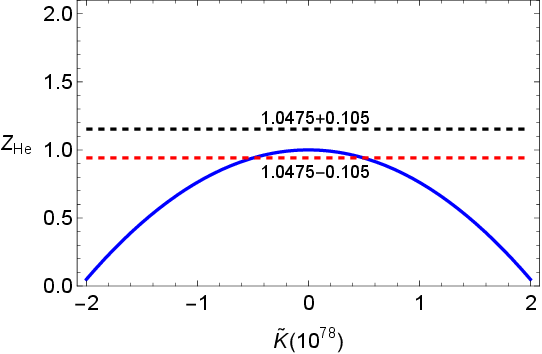}
\caption{$Z_{_{}^{4}\textrm{He}}$ vs $\tilde {K}$. The
experimental range is reported in (\ref {zhe}). We have fixed
$\eta_{10}=6$ and the temperature of freeze-out $T_f=1 MeV$.}
    \label{Fig1}
\end{figure}

\begin{figure} [H]
\includegraphics[scale=0.88]{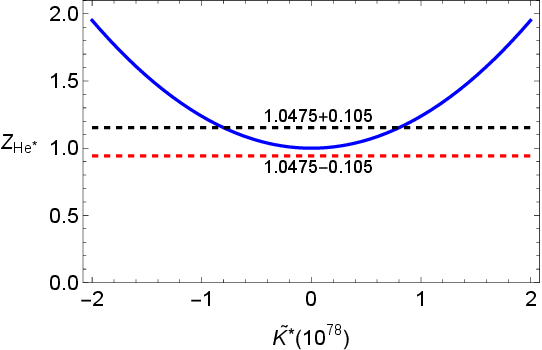}
\caption{$Z_{_{}^{4}\textrm{He}}$ vs $\Tilde{K^*}$. The
experimental range is reported in (\ref {zhe}). We have fixed
$\eta_{10}=6$ and the temperature of freeze-out $T_f=1 MeV$.}
    \label{Fig2}
\end{figure}

\begin{figure} [H]
\includegraphics[scale=0.86]{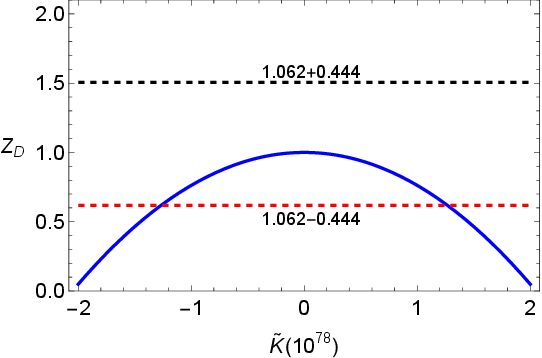}
\caption{ $Z_D$ vs $\tilde{K}$. The experimental range is reported
in (\ref {zobs}). We have fixed $\eta_{10}=6$ and the temperature
of freeze-out $T_f=1 MeV$.}
    \label{Fig3}
\end{figure}

\begin{figure} [H]
\includegraphics[scale=0.88]{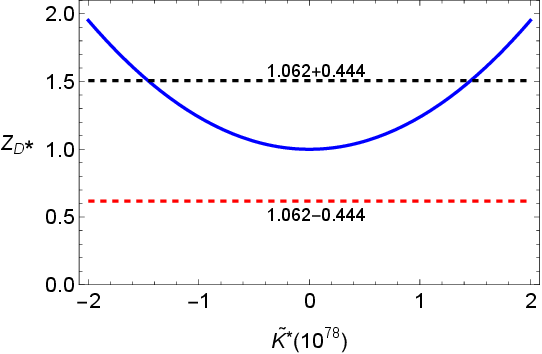}
\caption{$Z_D$ vs $\tilde{K^*}$. The experimental range is
reported in (\ref {zobs}). We have fixed $\eta_{10}=6$ and the
temperature of freeze-out $T_f=1 MeV$.}
    \label{Fig4}
\end{figure}

\begin{figure} [H]
\includegraphics[scale=0.88]{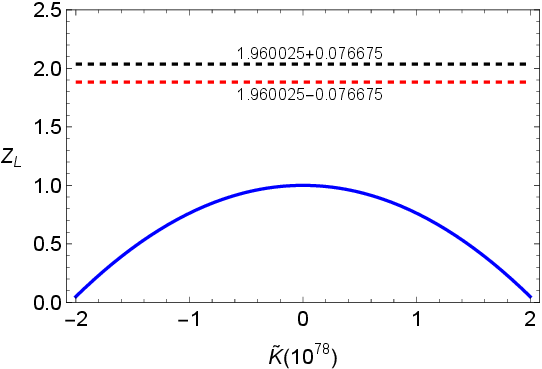}
\caption{$Z_{Li}$ vs $\tilde{K}$. The experimental range is
reported in (\ref {lit}). We have fixed $\eta_{10}=6$ and the
temperature of freeze-out $T_f=1 MeV$.}
    \label{Fig5}
\end{figure}

\begin{figure} [H]
\includegraphics[scale=0.88]{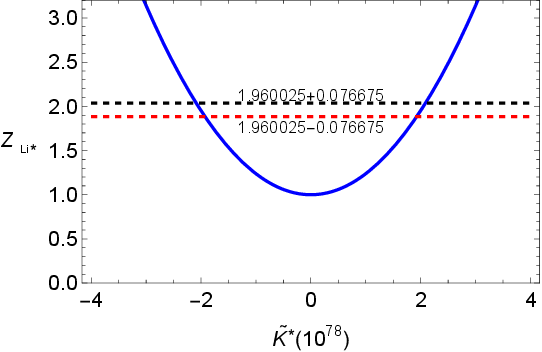}
\caption{$Z_{Li}$ vs $\tilde{K}^*$. The experimental range is
reported in (\ref {lit}). We have fixed $\eta_{10}=6$ and the
temperature of freeze-out $T_f=1MeV$.}
    \label{Fig6}
\end{figure}

Here we discuss the obtained results, deriving the bound on both
the Kaniadakis and dual Kaniadakis parameters. In Fig. \ref{Fig1}
we plot relation (\ref{zt}) taking into account the $-$ sign,
using the bounds presented in relation (\ref{zhe}). As we can see,
from the observational constraints on Helium abundance, the
rescaled Kaniadakis parameter $\tilde{K}=K/10^{-9}$ is constrained
to the range
\begin{equation} \label {dhes}
    -0.49 \times10^{-78}\lesssim \tilde{K} \lesssim 0.49 \times10^{-78}.
\end{equation}
The results for the dual Kaniadakis parameter are shown in the
Fig. \ref{Fig2}.  We observe that the dual Kaniadakis parameter
lies in the range
\begin{equation} \label {dhe}
{-0.8\times 10^{-78}\lesssim \tilde {K^*}\lesssim   0.8\times
10^{-78}.}
\end{equation}
We have shown the results for Deuterium (using Eq. (\ref {zobs})
and taking into account the $-$ sign in \eqref{zt}) in Fig.
\ref{Fig3}. We observe that the corresponding range of Kaniadakis
parameter $\tilde{K}$ is
\begin{equation} \label {deuts}
{-1.27 \times 10^{-78}\lesssim \tilde {K}\lesssim   1.27 \times
10^{-78}.}
\end{equation}
Let us note that the allowed  range of $\tilde{K}$ from Deuterium
(\ref {deuts}) is consistent with that obtained from
$_{}^{4}\textit{He}$ given by (\ref {dhes}).
The results for the dual Kaniadakis parameter are shown in the
Fig. \ref{Fig4}. It is seen that the dual Kaniadakis parameter
ranges as
\begin{equation} \label {deut}
   { -1.46 \times 10^{-78}\lesssim \tilde {K^*}\lesssim   1.46 \times 10^{-78}.}
\end{equation}
We observe that the allowed  range of $\tilde{K^*}$ from Deuterium
(\ref {deut}) is consistent with that obtained from
$_{}^{4}\textit{He}$ given by (\ref {dhe}).
In the case of Lithium, as we can see in Fig. \ref{Fig5}, the
results obtained from primordial Lithium abundance do not provide
a constrained range for the Kaniadakis parameter.
The results for the dual Kaniadakis parameter are shown in Fig.
\ref{Fig6}. The plot demonstrates that the range of $\tilde{K^*} $
for $_{}^{7}\textit{Li}$ overlaps with the constraints on
$\tilde{K^*}$ that $_{}^{4}\textit{He}$ and $_{}^{2}\textit{H}$
abundances provided in Eqs. (\ref {dhe}) and (\ref {deut}).
\begin{equation} \label {ZLi}
      {   -2.09\times^{-78}\lesssim \tilde {K^*} \lesssim 2.09 \times 10^{-78}.}
\end{equation}
Therefore, the allowed range for the dual Kaniadakis parameter,
based on the observed light elements abundances, is the
overlapping region of the intervals \eqref{dhe}, \eqref{deut} and
\eqref{ZLi}, and hence the acceptable values for the dual
Kaniadakis parameter are given by relation \eqref{dhe}. Consistent
with predictions, the deviation from standard Bekenestein-Hawking
expression is small.
\section{Time-Temperature relation in (Dual) Kaniadakis Cosmology}\label{tT}
In this section, we derive the relation between time and
temperature in the framework of (dual) Kaniadakis cosmology.
Modifications to the Friedmann equations influence the
thermodynamics of the early universe, leading to a different
relation between time and temperature compared to the standard
cosmological model. The state of thermal equilibrium in the early
universe leads to the conservation of entropy within a co-moving
volume \cite{Weinberg}
\begin{align}\label{sa}
s(T)a^3=\rm const.
\end{align}
Calculating the first derivative of the above equation with
respect to time results
\begin{align}
\dot{s}(T)a^3+3\dot{a}a^2s(T)=0 \mapsto
\frac{ds(T)}{dt}a^3=-3\dot{a}a^2s(T) \;.
\end{align}
Substituting \( \frac{\dot{a}}{a}=\left(
H_{GR}^2\pm\frac{\lambda}{H_{GR}^2} \right)^{1/2} \) into the
previous equation, we obtain
\begin{align}\label{3}
 \frac{ds(T)}{dt}=-3\left( H_{GR}^2\pm \frac{\lambda}{H_{GR}^2}
 \right)^{1/2}s(T),
 \end{align}
where can be reexpressed in the following manner
\begin{align}\label{dt}
dt=-\frac{ds(T)}{3s(T)}\left( H_{GR}^2\pm\frac{\lambda}{H_{GR}^2}
\right)^{-1/2}.
\end{align}
Integrating (\ref{dt}) with respect to temperature, the cosmic
time $t$ can be obtained as
\begin{align}\label{dis}
t=-\frac{1}{3}\int_{}^{} \frac{{s}'(T)}{s(T)}\left(
H_{GR}^2\pm\frac{\lambda}{H_{GR}^2}, \right)^{-1/2} dT,
\end{align}
where the prime symbol indicates a derivative with respect to the
temperature $T$.

Specifically, during any era in which the primary component of the
universe is a highly relativistic ideal gas, the entropy and
energy densities are given by the following equations
\cite{Weinberg}
\begin{eqnarray}\label{sT}
s(T)&=&\frac{2\mathcal{N}a_B T^3}{3},\\
 \rho(T)&=&\frac{\mathcal{N}a_B T^4}{2},\label{rhoT}
\end{eqnarray}
where $\mathcal{N}$ represents the total number of particles and
antiparticles, considering each spin state individually
\cite{Weinberg}. By expanding and substituting
$\frac{{s}'(T)}{s(T)}=\frac{3}{T}$, we derive the following result
\begin{align}\label{tT}
 t=-\int_{}^{}\frac{1}{TH_{GR}}\left( 1\pm\frac{\lambda}{2H_{GR}^4} \right)dT.
 \end{align}
Using $H^2_{GR}=\frac{8 \pi G }{3} \rho(T)$ as well as expression
$\rho(T)$ from Eq. (\ref{rhoT}), we reach
\begin{align}\label{3}
t=  -\int_{}^{}\left( \frac{1}{\Omega T^3}
\pm\frac{\lambda}{2\Omega^{5}T^{11}}\right)dT,
\end{align}
where we have defined $\Omega \equiv \sqrt{\frac{8\pi G\mathcal
{N} a_{B}}{6c^2}} $. Integrating yields
\begin{align}\label{tTbarrow}
t=\frac{1}{2\Omega T^2}\left( 1\pm\frac{\lambda}{10\Omega^4 T^8}
\right)+ \rm const.,
\end{align}
where $+$ and $-$ signs refer to the Kaniadakis entropy and the
dual Kaniadakis entropy, respectively. The above equation
describes how cosmic time $t$ is connected to temperature in the
context of (dual) Kaniadakis cosmology. In the limiting case where
$K=0$ we have $t=\frac{1}{T^2}\sqrt{\frac{3c^2}{16\pi
G\mathcal{N}a_{B}}} $ and GR is recovered \cite{Weinberg}.

We have used the modified Friedmann equation which was derived in
natural units where $\hbar=k_B=c=1 $. However, the right hand side
of Eq. \eqref{tTbarrow} must have the dimensions of time, hence if
we take into account all the constants in deriving the modified
Friedmann equations and include all physical constants properly to
ensure dimensional consistency, the constant $\lambda$ takes the
form
\begin{equation}\label{sil}
 \lambda= \frac{\pi ^2K^2}{2G^2}\mu^3\nu ,
 \end{equation}
 where we have defined
 \begin{equation}
 \mu\equiv {k_B c^3\over \hbar},\quad  \quad  \quad  \nu\equiv {\hbar c\over k_B}.
\end{equation}
As a result, the right side of Eq. \eqref {tTbarrow} is obtained
in terms of time. In the early stages of the universe, the
fundamental components included photons, three types of neutrinos,
three types of antineutrinos, along with electrons and positrons.
Under these conditions, the effective number of relativistic
degrees of freedom is given by $\mathcal {N} =43/4$. Thus, Eq.
\eqref{tTbarrow} yields, in cgs (centimeter-gram-second) units,
\begin{equation} \label{tTr}
t=0.994\;\left( {T\over10^{10 \; \circ} K} \right)^{-2}F(K,T)+\rm
const.,
\end{equation}
where we have defined
\begin{equation}
F(K,T)\equiv\left( 1\pm \frac{2.9474 \times10^{207}K^2}{T^8}
\right).
\end{equation}
Here the $+$ sign arises from the Kaniadakis entropy, while the
$-$ sign comes from dual Kaniadakis entropy. Again for $K=0$ we
have $F(K,T)\to1 $ and the result of GR is restored
\cite{Weinberg}.{As we mentioned earlier, in order to obtain the
correct numerical values in the time-temperature relation, it is
crucial to consider the fundamental constants in SI units.} {When
these fundamental constants are properly considered in the
definition of $\lambda$ (as in Eq. \eqref{sil}), energy density of
relativistic particles ($ \rho = \frac{\pi^2}{30} g_* \frac{k_B^4
T^4}{\hbar^3 c^3}$) and $H_{GR}=8\pi G/3c^2$, then Eq. \eqref{zt}
takes the form }

\begin{equation}
{Z(T)=1\pm\frac{2025}{384\pi^4}\frac{K^2 }{g_{*}^2G^4
T^8}\frac{c^{20}\hbar^4}{k_B^6}}.
\end{equation}
Using this revised relation, the Kaniadakis dual parameter is
normalized in conventional units to the value $K^*\simeq
10^{-64}$. Let us note that for $K\neq0$ and negative sign in $
F(K,T)$ (dual Kaniadakis), Eq. \eqref{tTr} admits an extremum, and
the time-temperature function in the dual Kaniadakis entropy
framework has a relative maximum at $T_{\rm max}(K)= \left( 1.4736
\times10^{208} \right)^{1/8}K^{1/4}$. Substituting the permissible
values of the dual Kaniadakis parameter into Time-Temperature
relation, the maximum temperature takes the value  $T_{max}\simeq
10^{10}$. This value is higher than the temperature required for
BBN $T_{max} \simeq 10^{9} $. Importantly, this maximum
temperature corresponds to an approximate time scale of
\(t(T_{max})\simeq 1\) second, which matches the time associated
with a temperature of \(10^{10}\,\mathrm{K}\)in the standard
cosmological model. This agreement further supports the
consistency of our approach within the established early-universe
framework and ensures that the universe in this model reaches
temperatures sufficient for baryogenesis, nucleosynthesis, and
standard matter particle creation, thus preserving viability
within the standard cosmological framework.

\begin{figure}[H]
\includegraphics[scale=0.88]{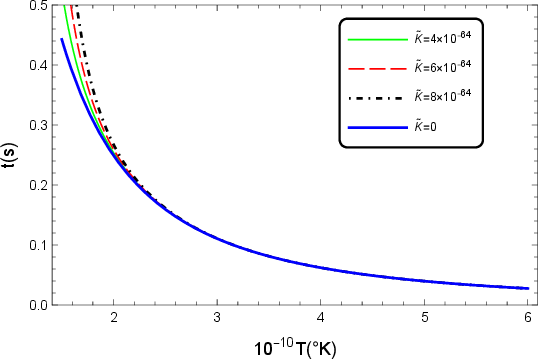}
\caption{The behavior of $T $ vs $t$ in
 Kaniadakis cosmology for early universe, and values of the Kaniadakis parameter that lie in the
 range{$\tilde{K}=[0, 4\times 10^{-64},6\times 10^{-64},8 \times 10^{-64}]$.}}
 \label{Fig7}
 \end{figure}

\begin{figure}[H]
\includegraphics[scale=0.88]{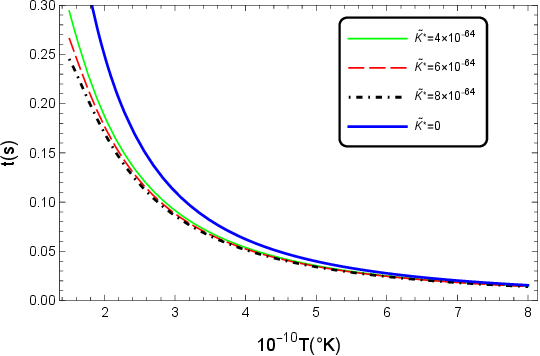}
\caption{The behavior of $T $ vs $t$ in dual Kaniadakis cosmology
for early universe, and values of the dual Kaniadakis parameter
that lie in the range {$\tilde{K}^{*}=[0, 4\times 10^{-64},6\times
10^{-64},8 \times 10^{-64}]$.}} \label{Fig8}
\end{figure}

In Fig. \ref{Fig7}, we plot the behavior of
temperature $T$ as a function of cosmic time $t$  for different
values of $\tilde{K}$.
We observe that as the Kaniadakis parameter
increases, the temperature of the early universe increases, too.
The results for the relation between time and temperature in dual
Kaniadakis cosmology are shown in Fig. \ref{Fig8}. It is seen that
as the dual Kaniadakis parameter increases, the temperature of the
early universe decreases as well.
It is also clear that the dual Kaniadakis entropy model does not
work well for the very early universe where $t\rightarrow 0$. The
physical reasons behind this behaviour of temperature in case of
the (dual) Kaniadakis may be due to the effects of non-extensive
thermodynamics. In standard cosmological model, both energy
density and entropy follow the usual extensive thermodynamics.
Kaniadakis cosmology represents non-extensivity, which leads to
more significant interactions and correlations between particles.
This results in an increased energy density for a given
temperature, effectively requiring a higher temperature to fit the
same energy density as in the GR case. The other reason can be the
modification of Friedmann equations which leads to modification in
expansion rate of the early universe. The larger values of
Kaniadakis parameter leads to a slower expansion rate at very
early universe, allowing more time for thermalization at higher
temperatures. This behavior is consistent with other non-extensive
thermodynamics frameworks (e. g., Tsalis cosmology), where
deformation parameters usually results in high-energy effects.
\section{Closing remarks \label{Closing}}
In this work, we have analyzed how Kaniadakis entropy and its dual
form influence gravitational interactions. By applying these
entropy measures within the entropic force framework, we first
obtained the modified form of the Newton's law of gravitation. We
then extended our analysis to the relativistic regime, by deriving
the modified Friedmann equations that govern cosmic evolution.
Furthermore, we derived the modified Friedmann equations by
establishing an equivalence between the difference in the degrees
of freedom on the boundary and in the bulk, and the variation of
the system's volume. This approach, based on the emergent gravity
scenario, incorporates entropy corrections and provides deeper
insights into cosmic evolution. Thus, by incorporating (dual)
Kaniadakis entropy, we have derived the modified Friedmann
equation through a different theoretical approach. The final
results are expressed in Eqs. (\ref{bkb}) and (\ref{bb}) where the
positive correction term originates from Kaniadakis entropy, while
the negative correction term arises from its dual formulation.


We have used BBN data in order to find the bounds on the free
parameter $K$ of (dual) Kaniadakis entropy. As we mentioned,
(dual) Kaniadakis entropy introduces additional terms into the
modified Friedmann equations. For consistency with BBN, these
corrections must be sufficiently small. In order to impose bounds
on the free parameters of the models, we explored the consequences
of (dual) Kaniadakis cosmology on the primordial light elements
formation. The analysis of light elements $_{}^{4}\textit{He}$,
$_{}^{2}\textit{H}$ and $ _{}^{7}\textit{Li}$ shows that the
allowed ranges obtained for the dual Kaniadakis parameter
$\tilde{K^*}$ derived from the primordial abundances of light
elements presented in relations (\ref {zhe}), (\ref {zobs}) and
(\ref {lit}), show significant overlap. Consequently, the
observationally allowed range for the dual Kaniadakis parameter,
constrained by primordial light element abundances, corresponds to
the intersection of the intervals given in  \eqref{dhe},
\eqref{deut} and \eqref{ZLi}, and hence the allowed values of dual
Kaniadakis parameter range as $ -0.8\times 10^{-78}\lesssim \tilde
{K^*}\lesssim   0.8\times 10^{-78}$. As expected, the deviations
from the standard cosmological model is small. Since the Lithium
problem refers to the discrepancy between the predicted and the
observed primordial Lithium abundance from BBN, if an allowed
range for the free parameter of the modified cosmological model
can simultaneously fit the observed abundances of
$_{}^{4}\textit{He}$, Deuterium (which are consistent with the
standard BBN model predictions) and Lithium, then it may be
possible to mitigate the Lithium problem without invoking
astrophysical corrections. This consistency between the allowed
ranges of the dual Kaniadakis parameter (obtained from the light
element abundances), may help to resolve the Lithium problem,
because it imposes strong constraints on the possible
modifications to standard cosmology. This suggests that the dual
Kaniadakis cosmology may potentially alleviate the Lithium
problem. On the other hand, the permissible range for the
Kaniadakis parameter, constrained by observational data on helium
and deuterium abundances, remains consistent. However, the
primordial Lithium abundance does not yield a valid range for the
Kaniadakis parameter.

We have also presented the formal relation between the cosmic time
$t$ and temperature of the universe in case of (dual) Kaniadakis
cosmology. We have plotted the behavior of temperature as a
function of cosmic time $t$, at the early stages of the universe,
and in the context of (dual) Kaniadakis cosmology. Our results
show that as the Kaniadakis parameter increases, the temperature
of the early universe increases, while an increase in the dual
Kaniadakis parameter leads to decreasing in the temperature of the
early universe. This behavior stems from non-extensive
thermodynamics, where stronger particle interactions increase
energy density at a given temperature, requiring higher
temperatures to fit GR predictions. Additionally, modified
Friedmann equations slow early-universe expansion rate, prolonging
high-temperature thermalization. Similar effects occur in other
non-extensive frameworks (e.g., Tsallis cosmology), where
deformation parameters lead to high-energy effects.
\acknowledgments{We are grateful to Shiraz university Research
Council. We also thank the anonymous referee for very insightful
and constructive comments, which helped us improve our paper
significantly. The work of A. Sheykhi is based upon research
funded by Iran National Science Foundation (INSF) under project
No. 4022705.}

\end{document}